\documentclass[twocolumn]{aastex63}
\usepackage{graphicx}	
\usepackage{amsmath}	
\usepackage{amssymb}	
\usepackage{scrextend}

\shorttitle{Decadal Timing of PSR J1544+4937}
\shortauthors{Kumari et al.}

\begin{document}
\title{Decade Long Timing Study of the Black Widow Millisecond Pulsar J1544+4937}
\correspondingauthor{Sangita Kumari}
\email{skumari@ncra.tifr.res.in}

\author[0000-0002-3764-9204]{Sangita Kumari}
\affiliation{National Centre for Radio Astrophysics, Tata Institute of Fundamental Research, S. P. Pune University Campus, Pune 411007, India}
\author[0000-0002-6287-6900]{Bhaswati Bhattacharyya}
\affiliation{National Centre for Radio Astrophysics, Tata Institute of Fundamental Research, S. P. Pune University Campus, Pune 411007, India}
\author[0000-0001-8801-9635]{Devojyoti Kansabanik}
\affiliation{National Centre for Radio Astrophysics, Tata Institute of Fundamental Research, S. P. Pune University Campus, Pune 411007, India}
\author[0000-0002-2892-8025]{Jayanta Roy}
\affiliation{National Centre for Radio Astrophysics, Tata Institute of Fundamental Research, S. P. Pune University Campus, Pune 411007, India}

\begin{abstract}
Results from 11 years of radio timing for eclipsing black widow millisecond pulsar (MSP) binary, J1544+4937, is presented in this paper. We report a phase-connected timing model for this MSP, using observations with the Giant Metrewave Radio Telescope (GMRT) at multiple frequencies and with Green Bank Telescope (GBT). This is the longest-duration timing study of any galactic field MSP with the GMRT. While extending the timing baseline from the existing 1.5 years to about a decade we report the first detection for a significant value of proper motion ($\mathrm{\mu_{T}} \sim$ 10.14(5) $\mathrm{mas/year}$) for this pulsar. Temporal variations of dispersion measure ($\mathrm{\Delta DM~ \sim 10^{-3}}$ pc $\mathrm{cm^{-3}}$) manifested by significant determination of 1st, 2nd, and 3rd order DM derivatives are observed along the line of sight to the pulsar. We also noticed frequency-dependent DM variations of the order of $\mathrm{10^{-3}~ pc~ cm^{-3}}$, which could arise due to spatial electron density variations in the interstellar medium. This study has revealed a secular variation of the orbital period for this MSP for the first time. We investigated possible causes and propose that variation in the gravitational quadrupole moment of the companion could be responsible for the observed temporal changes in the orbital period.
\end{abstract}

\keywords{pulsars: general; binaries: eclipsing, pulsars: individual }
\section{Introduction}
\label{sec:intro}
Out of 3200 known radio pulsars, about 16$\%$\footnote{\label{note1}\url{https://www.atnf.csiro.au/research/pulsar/psrcat/} as of 3rd August 2022} are millisecond pulsars (MSPs) with a spin period less than 30 ms. A majority of MSPs ($>$80\%) are found in binary systems, and are thought to evolve from normal pulsars by accreting matter from their companion \citep[e.g.][]{bhattacharya1991formation}. A special class of binary MSPs, called ``spider" MSPs, are in compact binary orbits with an orbital period of less than a day \citep{roberts2012surrounded}. Based on the mass of the companion ($\mathrm{m_{c}}$), spider MSPs are further divided into two categories: ``redback" (RB, 0.1M$_\odot<\mathrm{m_{c}}<$ 0.9M$_\odot$) and ``black widow" (BW, $\mathrm{m_{c}<0.05M_\odot}$). 

The population of spider MSPs has increased steadily in the last decade with the majority of them discovered via \emph{Fermi} directed surveys \citep[e.g.][]{ray2012radio,bhattacharyya2013gmrt,barr2013pulsar}. The possible reason could be the selection effect, where the energetic MSPs powering $\gamma$-ray sources are more likely to be in these types of interacting binary systems, or possibly that some of the $\gamma$-rays are coming from the interaction region itself \citep[e.g.][]{ray2012radio,roberts2012surrounded}. 

Although about 30$\%$\footref{note1} of the total MSP population is BW MSPs, long-term timing studies have been done only for a handful of such systems ($\sim$ 20\%\footref{note1} of total BW MSPs). 
Timing studies of BW MSPs are important in many ways. As the pulsar and the companion are in very close proximity in these systems, the highly energetic wind from the pulsar ablates the companion and complete evaporation of the companion could be one way to form isolated MSPs \citep[e.g.][]{polzin2020study}. Till now no BW MSP has been found where it is possible to ablate the companion within the Hubble timescale and a quest to find one such pulsar is still on \citep[e.g.][]{polzin2020study}. Also due to their compact binary orbits, the pulsars in such systems may accrete more mass from the companion compared to other MSPs during the low mass X-ray binary phase, which may result in a heavier neutron star \citep[e.g.][]{romani2022psr,romani2012psr,romani2021psr,van2011evidence}. The determination of the masses of BW MSPs may help to constrain the equation of state of the neutron star \citep[e.g.][]{antoniadis2013massive,fonseca2021refined}. 
Additionally, radio ephemeris obtained from such long-term studies of BW MSP systems aid the detection of gamma-ray pulsations and multi-wavelength follow-up investigations.

MSPs generally show stable timing because of their intrinsic rotational stability and small spin periods \citep{LorimerKramer}. On the other hand, BW MSPs are hard to time as the majority of such systems are susceptible to orbital period variation along with the variation of dispersion measure (DM) over long time scales. For example, PSR J2051$-$0827 \citep{shaifullah201621}; PSR J1959+2048 \citep{nice2000binary}; PSR J0023+0923 \citep{bak2020timing}; 47 Tuc J, 47 Tuc O \citep{freire2017long} and PSR J1731$-$1847 \citep{ng2014high} are some systems for which orbital period variations are reported.

PSR J1544+4937 is a BW MSP with a spin period of 2.16 ms discovered by the Giant Metrewave Radio Telescope  \citep[GMRT,][]{swarup1991asp} in a \emph{Fermi} directed search \citep{bhattacharyya2013gmrt}. Initial 1.5 years of timing analysis using the legacy GMRT system having 33 MHz bandwidth \citep{roy2010real} is reported by \cite{bhattacharyya2013gmrt}. This study found that PSR J1544+4937 is in a binary system with an orbital period of 2.9 hours orbiting around a low mass companion star (m$_{c} \ge 0.017$M$_\odot$). \cite{bhattacharyya2013gmrt} observed frequency-dependent eclipsing for this MSP with $\sim$ 13\% of the orbit eclipsed at $\sim$ 322 MHz. \cite{kansabanik2021unraveling} modeled the broadband radio spectrum in the optically thick-to-thin transition regime using multi-frequency wide bandwidth observations from the upgraded GMRT \citep{gupta2017upgraded} and identified synchrotron absorption by relativistic electrons as the possible cause for eclipsing.

\cite{tang2014identification} reported the optical identification of the companion to the MSP J1544$+$4937. This study inferred that the companion could either be a carbon, helium object, or a hydrogen brown dwarf based on the fitting of the light curves with different optical emitting models and distance to the pulsar.

The $\gamma$-ray pulsations from this MSP were discovered by \cite{bhattacharyya2013gmrt}, while using the radio timing ephemeris to fold the photons detected by the \emph{Fermi} Large Area Telescope (LAT).
The $\gamma$-ray detectability metric $\mathrm{\dot{E}^{\frac{1}{2}}/d^{2}}$ for PSR J1544+4937 is $\mathrm{7.4 \times 10^{16} ~erg^{\frac{1}{2}} kpc^{-2} s^{-\frac{1}{2}}}$, where $\mathrm{\dot{E}}$ is the spin-down luminosity and d is the distance to the pulsar. This value is comparable to other $\gamma$-ray detected MSPs \citep[Figure 12 of][]{abdo2010first} as well as other BW MSPs.

The decade-long timing of PSR J1544+4937 will aid in the studies of proper motion, dispersion measure, and orbital period variation. It will also allow us to investigate the possibility of inclusion of this system in the Pulsar Timing Array \citep[PTA, e.g.][]{detweiler1979pulsar,foster1990constructing}. The ephemeris from this timing study may also provide an improved detection significance in gamma-rays, subsequently allowing the high-energy studies of this system.

In this paper, we present 11 years of timing study for PSR J1544+4937. 
The details of the observations and the data analysis are discussed in Section \ref{sec:obs}. In Section \ref{sec:res} the results from 11 years of timing are presented. Section \ref{sec:Discussion} details the discussion and conclusion of this paper.

\begin{table*}[!htb]
\begin{center}
\caption{Summary of the observations.}
\label{tab:Table1}
\vspace{0.3cm}
\label{discovery}
\begin{tabular}{|l|l|l|l|l|l|l|}
\hline
Backend       & $T_{res}^{a}$ & $F_{res}^{b}$ & Bandwidth & S$_{min}$ & Span &  ${\sigma_{\tiny  TOA}}$ \\
&($\mu$s)&(MHz)&(MHz) &(mJy)&& ($\mu$s)\\
\hline
Legacy GMRT$^{\ast}$  & 61 & 0.064 & 33 & $0.05^{c}$            & 2011$-$2017 &   $2.5^{d}$          \\
\hline
GBT$^{\dagger}$  & 81 & 0.024 & 100 & 0.09$^{i}$            & 2012 &         3$^{j}$    \\
\hline
Upgraded GMRT$^{\ast\ast}$ & 81 & 0.048 & 200 & $0.02^{c}$               &  2018$-$2022 & $1^{d}$        \\
\hline
\end{tabular}
\end{center}
$^{\dagger}$ : \cite{guppi}\\
$^{\ast}$ : \cite{roy2010real}\\
$^{\ast\ast}$ : \cite{gupta2017upgraded} \\
$^{a}$: Time resolution\\
$^{b}$: Frequency resolution\\
$^{c}$: 5$\sigma$ detection sensitivity calculated using the radiometer equation \citep{LorimerKramer} considering  gain of 0.32 K/Jy for the Legacy GMRT system and 0.38 K/Jy for the upgraded GMRT system, system temperature $\sim$ 123 K including the sky temperature in the direction of J1544+4937, 20\% duty cycle, 25 antennas ($\sim $ number of antennas used in the observations), considering 2 polarisation's and 60 mins of observing time in coherent array mode.\\
$^{d}$: Error in TOA estimation calculated using the radiometer equation considering the above parameters for GMRT in 10 mins of observing time at 400 MHz.\\
$^{i}$: 5$\sigma$ detection sensitivity calculated using the radiometer equation \citep{LorimerKramer} considering  gain of 2 K/Jy, system temperature is around 65 K at the position of the pulsar including all effects, considering 2 polarisations and 60 mins of observing time.\\
$^{j}$: Error in TOA estimation calculated using the radiometer equation considering the above parameters for GBT in 10 mins of observing time at 350 MHz.\\

\vspace{1cm}
\end{table*}

\section{Observations and data analysis}
\label{sec:obs}
The majority of the observations reported in this paper were performed with the GMRT \citep{swarup1991asp,gupta2017upgraded}, which is a radio interferometric array consisting of 30 dishes, each 45 meters in diameter.
PSR J1544$+$4937 was localized to an accuracy of 5$''$  using continuum imaging with the full GMRT array followed by multi-pixel beamforming \citep{roy2012multi}. Such accurate localization allowed us to conduct sensitive follow-up observations using the coherent array mode apart from a few initial epochs after the discovery, which were performed in the incoherent array mode.

Initial observations from 2011$-$2017 were performed at the central frequency of 322 MHz and 607 MHz with the legacy GMRT \citep{roy2010real}, which is a 32 MHz bandwidth system. From 2017 onward the observations were taken with the upgraded GMRT \citep[uGMRT,][]{gupta2017upgraded}, at band 3 (300$-$500 MHz) and band 4 (550$-$750 MHz).
The simultaneous dual frequency observations were performed by splitting the total number of antennas into two sub-arrays at band 3 and band 4 for six observing epochs.
For three of the epochs observations were performed using the coherent de-dispersion technique. Coherent beam filterbank data at best achievable time-frequency resolution with the respective systems were recorded. 
In 2012, we used the Green Bank Telescope (GBT) to make several observations of PSR J1544$+$4937, in order to help determine the orbit of the system. These observations were taken using the GUPPI pulsar instrument \citep{guppi}, in 100\,MHz bandwidth modes centered at 350\,MHz.  One of the observations was made using coherent dedispersion and 10.24\,$\mu$s sampling with 512 frequency channels, giving excellent time resolution, while the others were made in an incoherent search mode with 81.92\,$\mu$s sampling and 4096 channels. The details of these observations are listed in Table \ref{tab:Table1}.

While processing the uGMRT data, we have used radio frequency interference (RFI) mitigation software, GMRT pulsar tool (gptool\footnote{\url{https://github.com/chowdhuryaditya/gptool}}) to mitigate the narrow-band and short duration broad-band RFI. Then we corrected the interstellar dispersion with the incoherent dedispersion technique \citep{LorimerKramer}. The dedispersed time series was folded with the known radio ephemeris for PSR J1544$+$4937 \citep{bhattacharyya2013gmrt} using $\it{prepfold}$ task of $\it{PRESTO}$ \citep{ransom2002fourier}.
The resultant mean pulse profile was cross-correlated with the template profile (high signal-to-noise profile from the previous observations). The temporal shift between these two profiles yielded the observed time of arrival (TOAs). The TOAs were generated using a {\it python} script {\it get\_TOAs.py} from $\it{PRESTO}$ \citep{ransom2002fourier}. We note that there is no significant frequency evolution of the pulse profile between band 3, band 4 for PSR J1544+4937. Calculated theoretical error on the TOAs ($\mathrm{\sigma_{TOA}}$) for different backends are presented in Table \ref{tab:Table1}. We have only considered the TOAs with uncertainties $<$ 9 $\mu$s. TOAs corresponding to the eclipse region (orbital phases $\sim$ 0.2 to 0.35) were removed from the analysis.

Using the timing model containing the information about the spin, astrometric and binary parameters of the pulsar (similar to one given in Table \ref{tab:Table2}) we can predict the arrival time of the pulses from the pulsar to the Earth. We compare these predicted TOAs with the observed TOAs. We have used $\it{tempo2}$ software package \citep{hobbs2006tempo2} to calculate the timing residual which is the difference between the observed and predicted TOAs.

We have investigated the timing properties of this system using two different orbital models, the ELL1 \citep{lange2001precision} and BTX model (D. Nice, unpublished). The ELL1 model is used for pulsars in low eccentric binary orbits, as in such systems the location of periastron does not show prominent features in the TOAs. On the other hand, the BTX model is the extension of BT model \citep{blandford1976arrival}, which allows fitting for higher order orbital frequency derivatives. We have fitted up to 5th order orbital frequency derivative (OFD) as beyond that successive OFDs were not determined with significance by $\it{tempo2}$ and fitting with these did not improve the timing residual further. This model also has no predictive power for the orbital period variations outside of the current TOA timeline.
  
\begin{figure}
   \centering
   \includegraphics[trim={0cm 0cm 0cm 0cm},clip,scale=0.3]{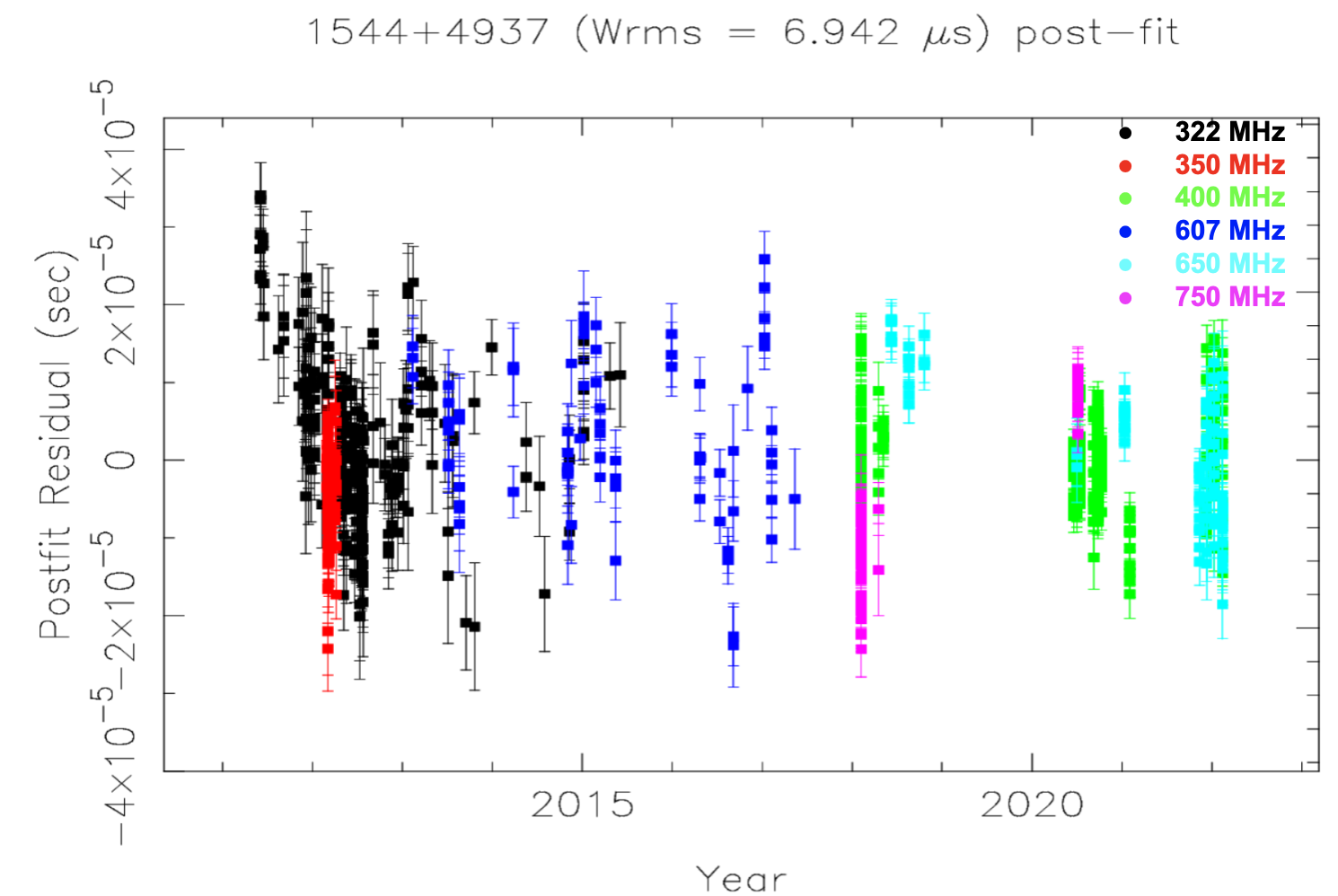} \includegraphics[trim={0cm 0cm 0cm 0cm},clip,scale=0.3]{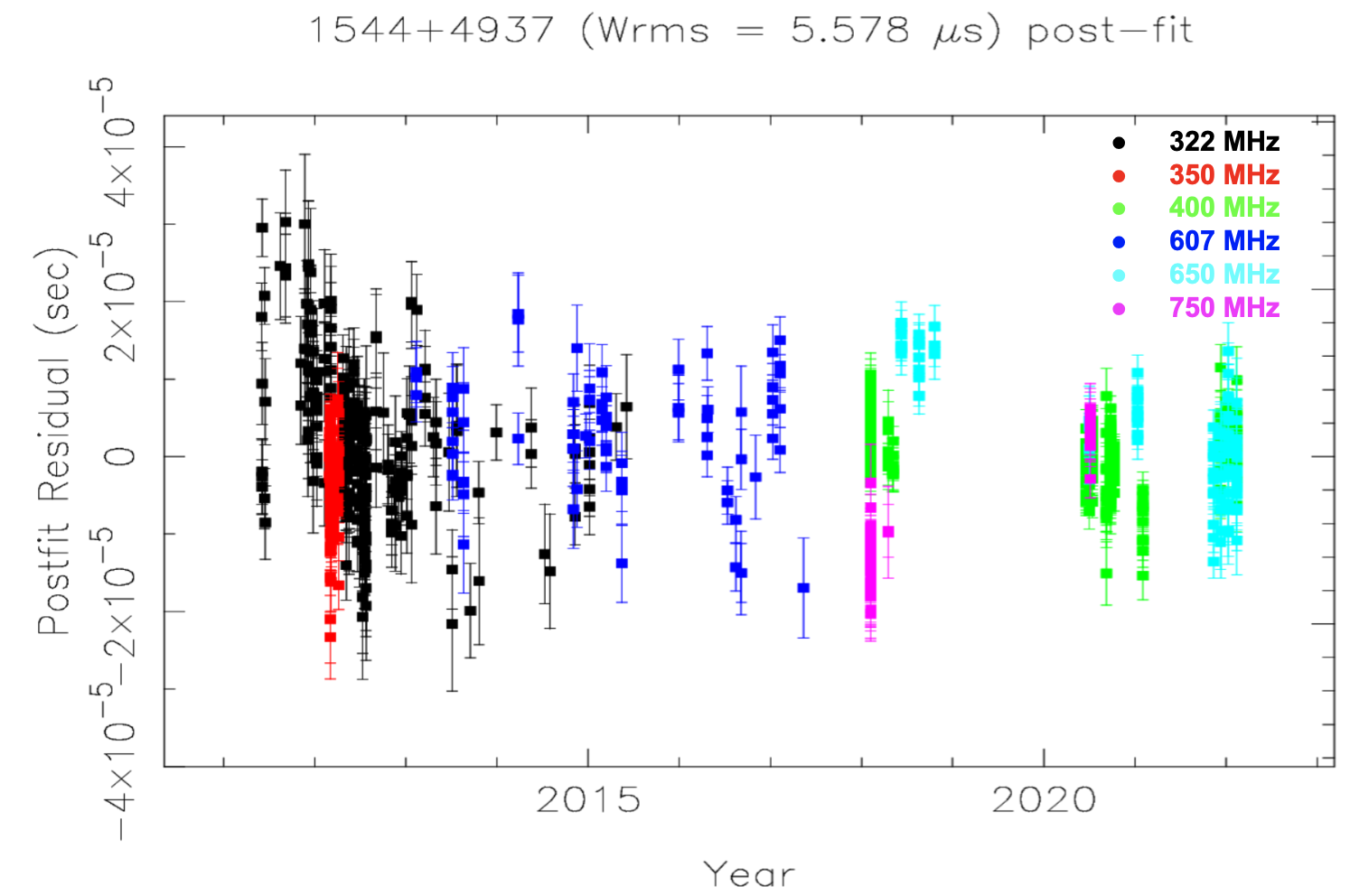}
    \caption{Post-fit timing residual from 11 years of timing for PSR J1544+4937 using the GMRT and the GBT observations. The TOAs generated for the GBT observations are shown in red. Upper panel: ELL1 model; Lower panel: BTX model.}
    \label{Figure:residual}
\end{figure}

The timing residual plots corresponding to these two models for the whole 11 year of timing span are given in Figure \ref{Figure:residual}, where the upper panel shows the timing residual corresponding to the ELL1 model and the lower panel shows the timing residual with the BTX model. Corresponding timing parameters are given in Table \ref{tab:Table2}. It is apparent from Figure \ref{Figure:residual}, that the BTX model better represents the orbital behavior of PSR J1544+4937 in comparison to the ELL1 model.

\begin{table*}
\centering
{\footnotesize
\caption{Timing parameters for PSR J1544+4937 using ELL1 and BTX model.
\label{tab:Table2}}
\begin{tabular}{|l|c||c|}
\hline
Model & ELL1 & BTX  \\
  \hline
 Timing Data Span     \dotfill           &55713.6$-$59622.2 & 55713.6$-$59622.2 \\
 Period epoch (MJD)\dotfill           & 55977.0 & 55977.0 \\
 Total time span (year) \dotfill                                & 10.7                         & 10.7              \\
 Number of TOAs\dotfill  &866 & 866 \\
 Reduced Chi-square \dotfill       &  7.5065       &  4.8642\\
 Post-fit residual rms ($\mu$s)\dotfill       & 6.942  & 5.578  \\
 \hline
 Right ascension (J2000)\dotfill &
 15$^\mathrm{h}$44$^\mathrm{m}$04\fs4911(1) &
 15$^\mathrm{h}$44$^\mathrm{m}$04\fs49135(9) \\
 Declination (J2000)\dotfill     &
 $+$49\degr37\arcmin55\farcs374(1) &
 $+$49\degr37\arcmin55\farcs3748(9)  \\
 Proper motion in RA (mas year$^{-1}$)\dotfill &$-$4.46(5) & $-$4.54(4) \\
 Proper motion in DEC (mas year$^{-1}$)\dotfill &$-$9.15(6) & $-$9.06(5) \\
 Spin frequency $f$ (Hz)\dotfill &
 463.11553585637(2) & 463.11553585640(1)\\
 Spin frequency derivative $\mathrm{\dot{f}}$ (Hz s$\mathrm{^{-1}}$)\dotfill &
 $-$5.998(1)$\times$10$^{-16}$  &
 $-$5.9996(9)$\times$10$^{-16}$ \\
Dispersion measure $\mbox{DM}$ (pc~cm$^{-3}$) &  23.2246(1)       & 23.22442(9) \\
 DM $1^{\mathrm st}$derivative $\mbox{DM1}$ (pc~cm$^{-3}$ year$^{-1}$) \dotfill & 0.00095(5)    & 0.00115(4)\\
DM $2^{\mathrm nd}$ derivative $\mbox{DM2}$ (pc~cm$^{-3}$ year$^{-2}$)\dotfill & $-$ 0.00041(2)   & $-$ 0.00048(1)\\
DM $3^{\mathrm nd}$ derivative $\mbox{DM3}$ (pc~cm$^{-3}$ year$^{-3}$)\dotfill & 5.9(4) $\times$ $10^{-5}$   & 7.08(3) $\times$ $10^{-5}$\\
 Orbital period $\mathrm{P_{b}}$ (days)\dotfill & 0.12077298998(8)  &  \\
 Orbital frequency FB0 \dotfill & & 9.58332993(2) $\times$ $10^{-5}$\\
 Orbital period 1st derivative (s/s)\dotfill & 6.6(4) $\times$ $10^{-13}$ & \\
 Orbital frequency $1^{\mathrm st}$derivative FB1 \dotfill & $-$ & 1.7(1)$\times$ $10^{-19}$  \\
 Orbital frequency $2^{\mathrm st}$derivative FB2 \dotfill & $-$ &-1.26(9)$\times$ $10^{-26}$ \\
 Orbital frequency $3^{\mathrm st}$derivative FB3 \dotfill & $-$ & 3.9(3)$\times 10^{-34}$\\
 Orbital frequency $4^{\mathrm st}$derivative FB4 \dotfill & $-$ & -6.3(5)$\times 10^{-42}$ \\
 Orbital frequency $5^{\mathrm st}$derivative FB5 \dotfill & $-$ & 4.4(3) $\times 10^{-50}$\\
 Projected semi-major axis $\mathrm{x}$ (lt-s) & 0.0328665(4) & 0.0328656(3) \\
 Time of ascending node $\mathrm{T_{ASC}}$ (MJD)\dotfill & 56124.7700998(4) & $-$\\
 $\mathrm{\epsilon_1 (e\cos{\omega})}$ \dotfill & 0.00003(2) & $-$  \\
 $\mathrm{\epsilon_2 (e\sin{\omega}})$ \dotfill & 0.00009(2) & $-$  \\
Eccentricity (e) \dotfill & 0.00010 & 5(1)$\times 10^{-5}$ \\ 
Epoch of periastron $\mathrm{T_{o}}$ (MJD) \dotfill & 56124.7765260 & 56124.7701015(4) \\
Longitude of periastron,  $\mathrm{\omega}$ (\degr) \dotfill & 19.15  & 0\\
 \hline
 \multicolumn{3}{c}{Derived parameters} \\
  \hline
 Galactic longitude, $\mathrm{l}$ (\degr) \dotfill    &  207.65 & 207.65 \\
 Galactic latitude, $\mathrm{b}$ (\degr) \dotfill    & 65.88 & 65.88  \\
 DM distance (kpc) \dotfill              & 1.2 $^\dagger$                        & 1.2$^\dagger$ \\
 & 2.9$^\ddagger$ & 2.9$^\ddagger$\\

 Observed $\mathrm{\dot{P}}$, $\mathrm{\dot{P}_{\rm obs}}$ ($\mathrm{s\, s^{-1}}$) \dotfill                             & 2.7966(5) $\times 10^{-21}$& $2.79735(4)\times 10^{-21}$  \\
 Intrinsic $\mathrm{\dot{P}}$, $\mathrm{\dot{P}_{\rm int}}$ ($\mathrm{s\,s^{-1}}$) \dotfill                             & 8.1 $\times 10^{-22}$ & $ 8.1 \times 10^{-22}$  \\
 Energy loss rate $\mathrm{\dot{E}}$ ($10^{33} \rm \, erg \, s^{-1}$)\dotfill                   &  12       &  12 \\
 Characteristic age, $\mathrm{\tau_c}$ (Gyear)\dotfill                &  42         & 42 \\
 Surface magnetic field, $\mathrm{B_0}$ (Gauss)\dotfill      &  $4.2 \times 10^{7}$         & $4.2 \times 10^{7}$ \\
 Total proper motion, $\mathrm{\mu_{T}}$ (mas year$^{-1}$) \dotfill       & 10.18(6)     & 10.14(5)  \\
 Companion mass, $\mathrm{M_c}$  (M$_{\odot}$)      \dotfill                           & 0.01697$<$ 0.0196 $<$ 0.0393 &  0.01697 $<$ 0.0196 $<$ 0.0393\\\hline
\end{tabular}
}\\
{\small $^\ast$ Errors correspond to 1$\sigma$.\\
$^\dagger$ using the NE2001 model electron density distribution\\
$^\ddagger$ using the YMW16 model of electron distribution\\
We note that the calculated DM distance is model dependent.\\
Timing uses DE405 solar system ephemeris.\\
The numbers in the parenthesis are uncertainties in preceding digits.}
\end{table*}

\section{Results}
\label{sec:res}
\subsection{New parameter estimation}
A decade of timing presented in this paper allowed the estimation of new parameters which were not detected by \cite{bhattacharyya2013gmrt}. It includes significant detection of DM derivatives, proper motion, higher order orbital period derivatives, and eccentricity for PSR J1544+4937 using the BTX model. 
Additionally, folding the Fermi-LAT gamma-ray photons with the radio ephemeris obtained from this study resulted in an improved detection \citep[$\sim$ 10$\sigma$ as compared to a 5$\sigma$ detection reported in][]{bhattacharyya2013gmrt}. However, a similar detection was already achieved using follow-up observations with the Nan\c{c}ay telescope combined with timing LAT photons and will be part of the Third Fermi-LAT Pulsar Catalog (LAT Collaboration 2023, in prep.).

\subsubsection{Proper motion}
Studies of proper motion are of paramount importance to constrain various models explaining the formation of the MSPs \citep[e.g.][]{tauris1996origin}. 
Knowledge of the proper motion of the MSPs provides more accurate values of spin-down rates ($\mathrm{\dot{P}}$) assuring that pulsars find their proper place in the $\mathrm{P-\dot{P}}$ diagram. Correcting for the spin-down rates also provides the improved value of magnetic field \citep[proportional to $\mathrm{\sqrt{P\dot{P}}},$][]{LorimerKramer} and characteristic age \citep[proportional to $\mathrm{P/\dot{P}},$][]{LorimerKramer} of the pulsar.

Total proper motion is defined as $\mathrm{\mu_{T} =  \sqrt{\mu_{\alpha}^2+\mu_{\delta}^2}}$ where $\mathrm{\mu_{\alpha}}$ is the proper motion in right ascension (RA) and $\mathrm{\mu_{\delta}}$ is the value of proper motion in declination (DEC).
In this study we have estimated $\mathrm{\mu_{T}}$ of 10.14(5) mas/year.
We have calculated the 2$-$D transverse velocity ($\mathrm{V_{T}}$) using the relation $\mathrm{V_{T}=4.74~\mu_{T}(mas/year)d(kpc)}$, where $d$ is the distance to the pulsar.
Using two galactic electron density models NE2001 \citep{cordes2002ne2001} and YMW16 \citep{yao2017new}, the DM distance ($d$) is estimated as 1.2 kpc and 2.9 kpc respectively.  
We estimated the $\mathrm{V_{T}}$ to be 140 km/s (YMW16) and 58 km/s (NE2001). 

For the PSR J1544+4937 we calculated the intrinsic spin period derivative to be $\mathrm{8.1 \times 10^{-22}}$ s/s, using equation (2) of \cite{toscano1999millisecond} considering $\mathrm{V_{T}}$ of 140 km/s. We have also determined the characteristic age and magnetic field for PSR J1544+4937 using the corrected spin period derivative as listed in Table \ref{tab:Table2}.

\subsubsection{Variations in dispersion measure}
DM is a measure of the integrated electron density along the line of sight to the pulsar. It gets affected by the motion of the pulsar or by the dynamical evolution of ISM. Under the assumption $\mathrm{f_{p}=8.98 \, kHz \times \sqrt{\frac{n_{e}}{cm^{-3}}}<< f}$ and $\mathrm{f_{c}=2.80 \, MHz \times\frac{B}{1G}<<f}$, where $\mathrm{f_{p}}$ is the plasma frequency, $\mathrm{f_{c}}$ is the cyclotron frequency and $\mathrm{f}$ is the observing frequency, the time delay of the radio signal at frequency $\mathrm{f}$ with respect to a signal of infinite frequency is given as \citep{LorimerKramer},
\begin{equation}
\label{eqn:dispersion law}
\mathrm{t = \frac{D \times DM}{f^{2}} }  
\end{equation}
where D is the dispersion constant $\sim$ $4.41 \times 10^{3}$ $\mathrm{MHz^{2}pc^{-1}cm^{-3}s}$, t is in seconds, f is in MHz and DM is in pc $\mathrm{cm^{-3}}$.

Time delay in equation (\ref{eqn:dispersion law}) is inversely proportional to the square of the observing frequency. Hence at low frequencies, even a small change in DM will produce large time delays and are easily detectable. Such variations can only be precisely measured and differentiated from other noise sources when simultaneous dual frequency or wide bandwidth data is available.

For observations from 2011 to 2017 taken with the legacy GMRT system at 322 and 610 MHz, the DM values were estimated using the near dual frequency epochs which were separated by a few days. For sensitive observations with the uGMRT system available between 2017 to 2022, the DM values were determined using simultaneous multi-frequency observations having 200 MHz bandwidth centered at 400 and 650 MHz. Additionally, we estimated DM for the uGMRT band 3 observations using the sub-band TOAs, where the observing bandwidth is divided into four sub-bands.

\begin{figure}[!htb]
\centering
\includegraphics[trim={0cm 0cm 0cm 0cm},clip,scale=0.245]{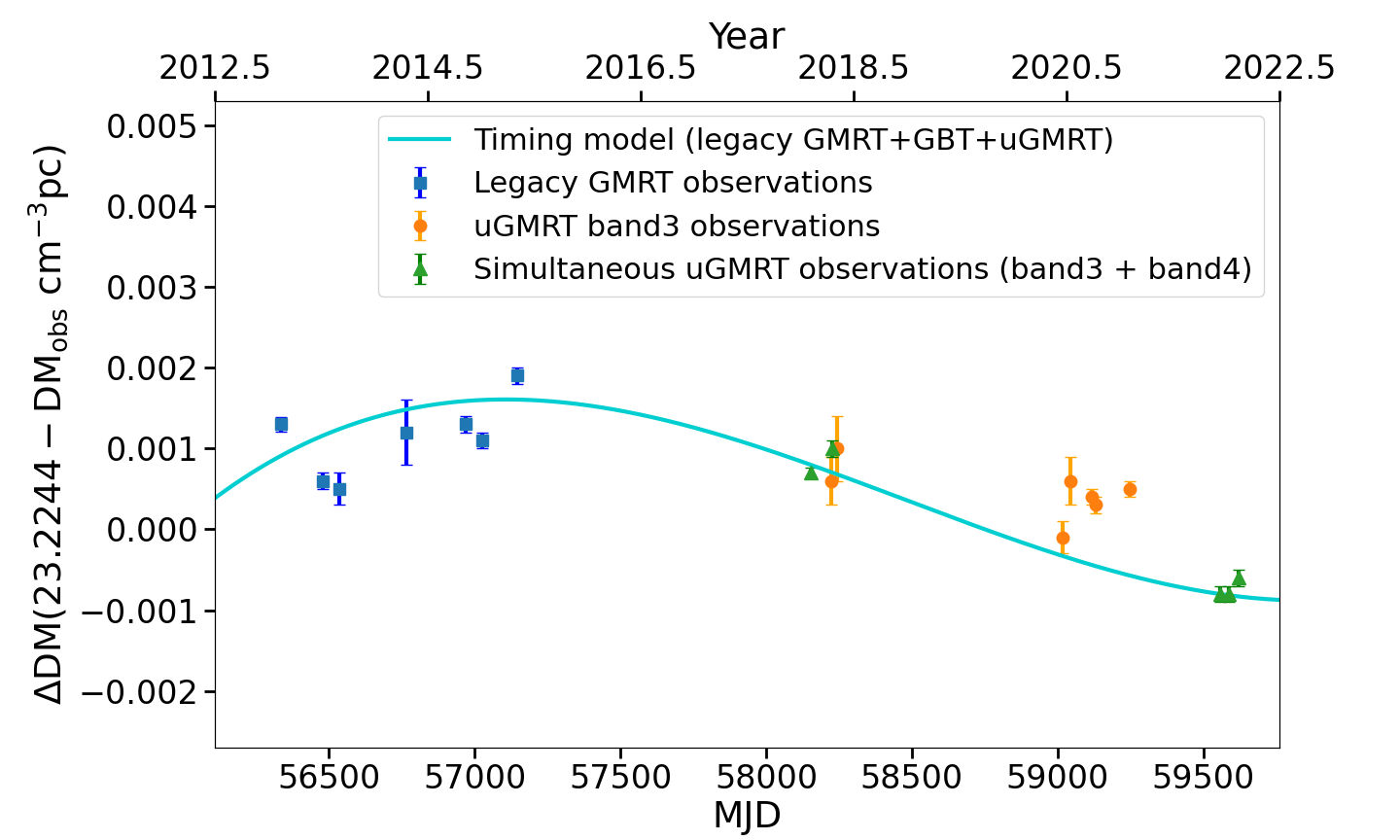}
\caption{Variation of DM with time. The blue square points represent the DM values determined using nearby dual frequency epochs (separated by a few days) with the 32 MHz legacy GMRT system. The orange circular points represent the estimated DM values (DM precision $ \sim 10^{-4}$ pc $\mathrm{cm^{-3}}$) using sub-band TOAs for the uGMRT observations at 400 MHz (having a bandwidth of 200 MHz). The green triangular points represent the DM determination using simultaneous dual frequency observations at band 3 and band 4 (DM precision $\sim 10^{-5}$ pc $\mathrm{cm^{-3}}$). The aqua blue line shows the $\it{tempo2}$ fit using a model with 3 DM derivatives.}
\label{Figure:longtermDMvariation}
\end{figure}

We have fitted up to third-order DM derivatives in the timing model as listed in Table \ref{tab:Table2}. Figure \ref{Figure:longtermDMvariation} shows long-term DM variation for PSR J1544+4937 for observations spanning from 2012 to 2022. It can be seen that the DM derivatives from the timing model provide a reasonable fit to the observed DM variation. Moreover, fitting the observed DM variations has allowed us to time PSR J1544+4937 more precisely.
 
\begin{figure}
\centering
\includegraphics[trim={0cm 0cm 0cm 0cm},clip,scale=0.24]{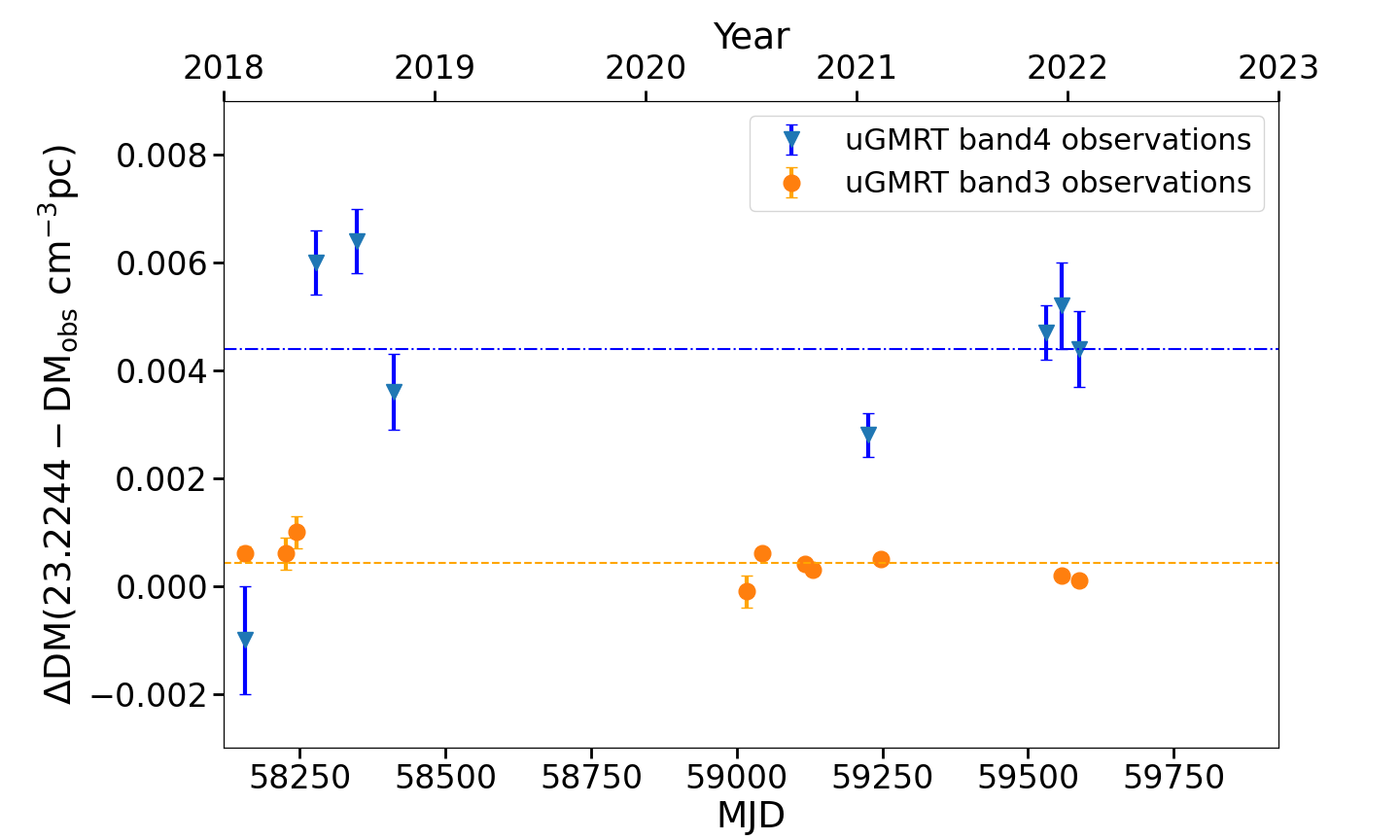}
\caption{Frequency-dependent DM variation observed for PSR J1544+4937. The orange circular points represent the DM values estimated using sub-band TOAs with the uGMRT observations at band 3 and the blue inverted triangular points represent the DM values estimated using sub-band TOAs with the uGMRT observations at band 4. The blue dashed dot line (-.) is the median value of the  $\mathrm{\Delta DM}$ for band 4 and the orange dashed line (-\--)} is the median value of the $\mathrm{\Delta DM}$ for band 3. The frequency-dependent DM variation between band 3 to band 4 is of the order of $10^{-3}$ pc $\mathrm{cm^{-3}}$.
\label{Figure:FrequencydependentDM}
\end{figure}

Along with the long-term temporal DM variations we have also observed frequency-dependent DM variations for PSR J1544+4937. To study the frequency-dependent variations, the DM values at band 3 and band 4 were determined using sub-band TOAs where the observing bandwidth was divided into four sub-bands for each band. Figure \ref{Figure:FrequencydependentDM} shows that observed variation between the two bands is of the order of 10$^{-3}$ pc $\mathrm{cm^{-3}}$, where DM estimated in band 3 is generally lower than those in band 4. 
 
\subsubsection{Orbital period variations}
We studied the orbital parameter variability for PSR J1544+4937, using the BTX model. We have fitted up to the 5th order orbital frequency derivative since beyond this successive orbital frequency derivatives did not affect the reduced $\chi^{2}$ and were not determined with significance. 

 The phase of the orbit ($\mathrm{\phi}$) according to the BTX model at a time (t) is given as,
 \begin{equation}
\mathrm{\phi_{BTX}(t)=\phi(t_{o})+n_{b}(t-t_{o})+\dot{n}_{b}\frac{(t-t_{o})^{2}}{2\,!}+\ddot{n}_{b}\frac{(t-t_{o})^{3}}{3\,!}+...}
\label{eqn:BTX}
\end{equation}
where $\mathrm{\phi (t_{o})}$ is the orbital phase at a reference time $\mathrm{t_{o}}$, $\mathrm{n_{b}}$ is the orbital frequency and $\mathrm{\dot{n}_{b}}$ , $\mathrm{\ddot{n}_{b}}$ are higher order orbital frequency derivatives at $\mathrm{t_{o}}$.

If there were no change in the orbital period with time (i.e. higher order orbital frequency derivatives are zero), then equation (\ref{eqn:BTX}) gives the orbital phase at any time t as predicted by a simple Keplerian model,
\begin{equation}
\mathrm{\phi_{Keplerian}(t)=\phi(t_{o})+n_{b}(t-t_{o})}
\label{eqn:keplerian}
\end{equation}

Subtracting the equation (\ref{eqn:keplerian}) from (\ref{eqn:BTX}), we get,
\begin{equation}
\mathrm{\Delta \phi =\dot{n}_{b}\frac{(t-t_{o})^{2}}{2\,!}+\ddot{n}_{b}\frac{(t-t_{o})^{3}}{3\,!}+.....}
\end{equation}
 or,
\begin{equation}
\mathrm{\Delta \phi = \sum_{k=1}^{5} n_{b}^{k}\frac{(t-t_{o})^{k+1}}{k+1 \,!}}
\end{equation}
where k goes from 1 to 5, since we are fitting up to 5 orbital frequency derivatives.

Also, $\mathrm{\Delta T_{o}}$ and $\mathrm{\Delta \phi} $ are related as \citep{ng2014high},
\begin{equation}
\label{eqn: DeltaTo}
\mathrm{\Delta T_{o} = \Delta \phi \times P_{b} }  
\end{equation}
where $\mathrm{P_{b}}$ is the orbital period.

\begin{figure}
   \centering
   \includegraphics[trim={1.5cm 0cm 1.9cm 0cm},clip,scale=0.22]{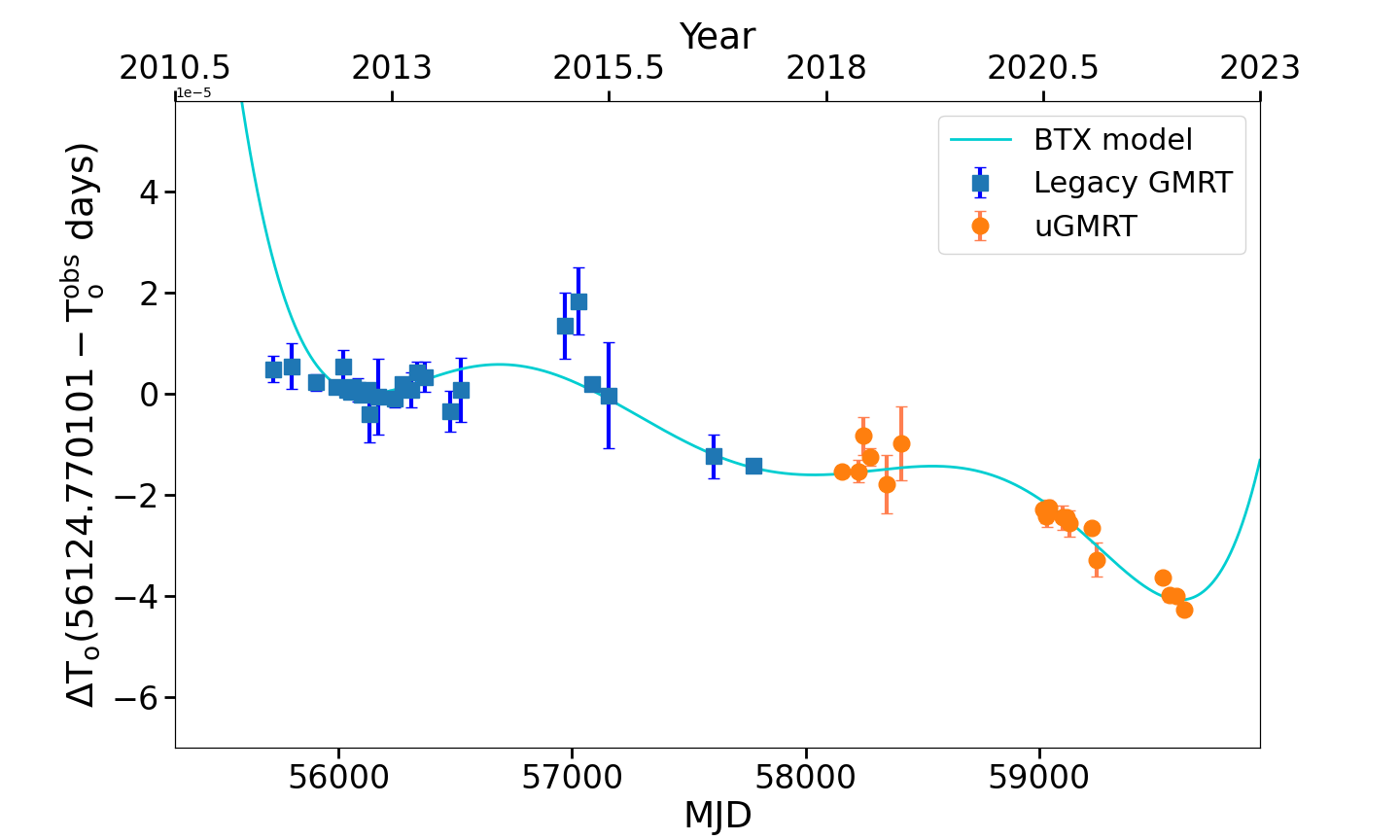} 
   \includegraphics[trim={0cm 0cm 1.0cm 0cm},clip,scale=0.225]{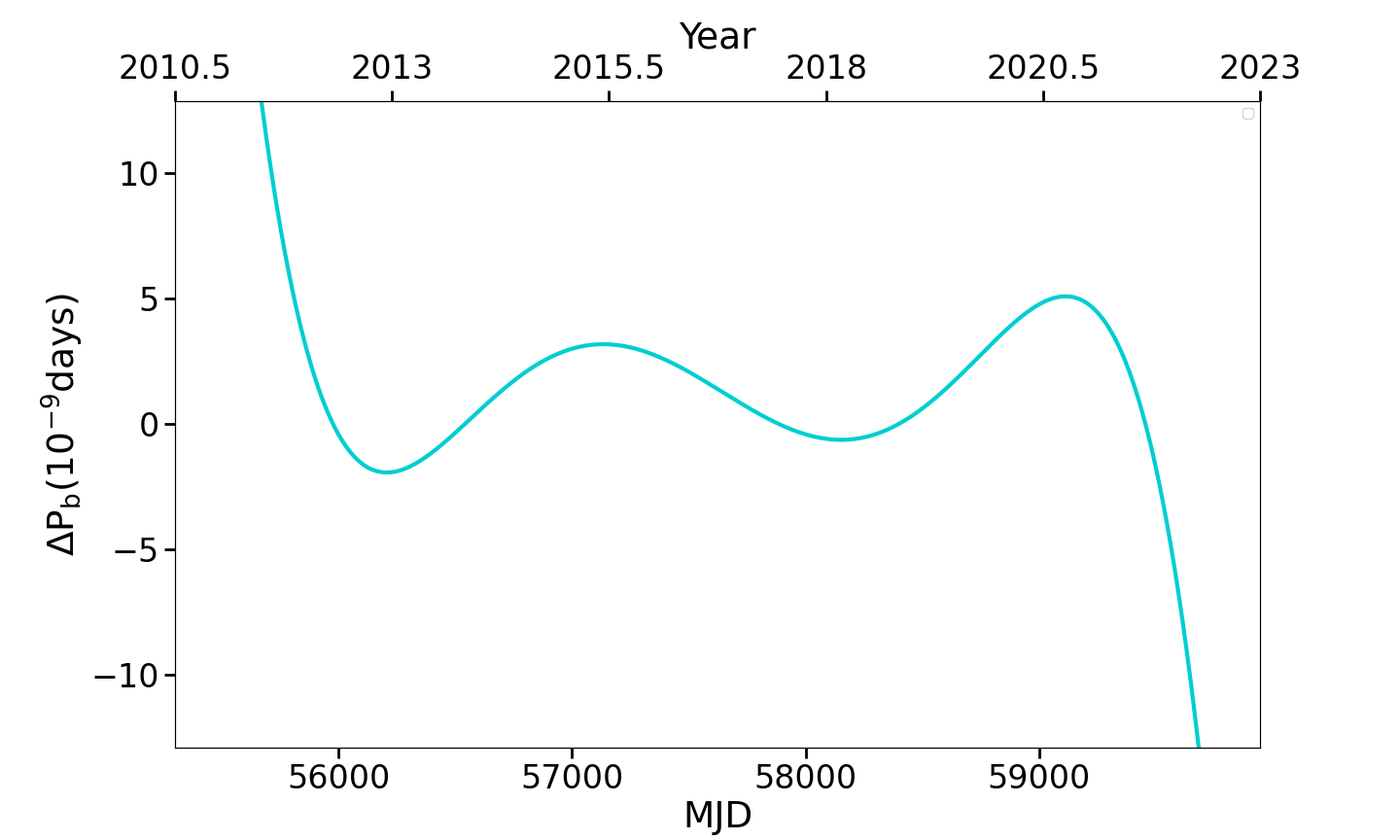}
    \caption{Upper panel: Variation of the epoch of periastron ($\mathrm{T_{o}}$) for PSR J1544+4937. The aqua blue line shows the variation of $\mathrm{T_{o}}$ as predicted by the BTX model. Blue square and orange circular points show the $\mathrm{T_{o}}$ values corresponding to individual epochs determined using the legacy GMRT and uGMRT data. Lower panel: Variation of the orbital period for PSR J1544+4937 as predicted by the BTX model shown by the aqua blue line.}
    \label{Figure:To_variation}
\end{figure}

Prediction by the BTX model according to equation (\ref{eqn: DeltaTo}) is shown by the aqua blue curve in the upper panel of Figure \ref{Figure:To_variation}. 
To confirm this prediction, we followed the procedure described below. 

For the observations with the legacy GMRT system from 2011$-$2017, we divided the data sets into smaller chunks, where each chunk consisted of around 6$-$8 TOAs separated by a few days. Whereas the observations with the uGMRT system spanning from 2018$-$2022 were sensitive enough to create around 10$-$20 TOAs for each epoch. 

Starting with the BTX model (obtained from 11 years of timing), we froze all the parameters except $\mathrm{T_{o}}$ and kept the value of all higher-order orbital frequency derivatives as zero. Then for each chunk (for legacy GMRT) and epoch (for uGMRT) we fitted only for $\mathrm{T_{o}}$ using $\it{tempo2}$ and determined the corresponding  $\mathrm{T_{o}^{obs}}$. From Figure \ref{Figure:To_variation}(a) it can be seen that the BTX model prediction (presented by aqua blue curve) is in close agreement with the observed variations of the epoch of ascending node (presented by blue square and orange circular points for legacy GMRT and uGMRT systems respectively).
The corresponding predicted orbital period variation by the BTX model is shown in the lower panel of Figure \ref{Figure:To_variation} by the aqua blue curve. 


\section{Discussion and Conclusions}
\label{sec:Discussion}
\subsection{Proper motion}
We report the first detection of proper motion and the corresponding $\mathrm{V_T}$ $\sim$ 140 km/s for PSR J1544+4937. We determined the mean $\mathrm{V_T}$ for a sample of BW MSPs as well as for a sample of other MSPs, whose proper motion and distances were available in the ATNF\footnote{https://www.atnf.csiro.au/research/pulsar/psrcat/} pulsar catalogue.
We have found mean $\mathrm{V_T}$ $\sim$ 147 km/s for a sample of 25 BW MSPs. For a sample of 59 MSPs (other than BW MSPs), mean $\mathrm{V_T}$ is $\sim$ 113 km/s. Observed higher value of mean $\mathrm{V_T}$ for the BW MSPs compared to the other MSPs may be biased by small number statistics. Considering different evolutionary mechanisms for the MSP formation, \cite{tauris1996origin} predicted mean recoil velocity for MSPs to be $\sim$ 110 km/s assuming symmetric supernova explosion and $\sim$ 160 km/s with the inclusion of random asymmetric kicks.

\cite{tauris1996origin} also predicted an anti-correlation between the orbital period and the recoil velocity. Based on this prediction, MSPs in the tight binary orbit (BW MSPs) should have high recoil velocities compared to other MSPs. We also found that the mean $\mathrm{V_T}$ obtained for the BW MSPs (147 km/s) is higher than the other MSPs (113 km/s). However, we note that possible errors on the value of $\mathrm{V_T}$ contributed by uncertainty in the DM distances estimated by the YMW16 model remain unaccounted for.

Previous determination of mean $\mathrm{V_T}$ with DM distance estimated from NE2001 Galactic electron density model reported somewhat lower values. For example, \cite{toscano1999millisecond} reported a mean $\mathrm{V_T}$ of 85$\pm$13 km/s for a sample of 23 MSPs. Further studies by \cite{hobbs2005statistical} and \cite{desvignes2016high} reported a mean $\mathrm{V_T}$ of 87$\pm$13 km/s and 92$\pm$10 km/s respectively for a sample of 35 and 76 MSPs. 

\cite{hobbs2005statistical} and \cite{desvignes2016high}  also concluded that the mean $\mathrm{V_T}$ for isolated pulsars is comparable with that for the binary MSPs. We have also calculated the mean $\mathrm{V_T} \sim 122$ km/s for a sample of 32 isolated MSPs. The mean $\mathrm{V_T}$ for BW MSPs is relatively higher than the mean $\mathrm{V_T}$ of isolated MSPs in our sample. 
This slight difference in the mean $\mathrm{V_{T}}$ between the two populations, may indicate that the BW MSPs may not be the progenitors of the isolated MSPs. Although a larger sample of isolated and BW MSPs is required to come to any conclusion. From the eclipse study also, so far no BW MSP system is found where the mass loss rate ($\mathrm{\dot{m}_{c}}$) is sufficient to evaporate the companion completely within the Hubble time scale. For example, for PSR J2051$-$0827, $\mathrm{\dot{m}_{c}=10^{-14}}$ M$_\odot$/year \citep{stappers1996probing}; for PSR J1959+2048 and PSR J1816$+$4510, $\mathrm{\dot{m}_{c}=10^{-12}}$ and $\mathrm{2 \times 10^{-13}}$ M$_\odot$/year \citep{polzin2020study} respectively. For PSR J1544+4937, from the eclipse study, we obtained  $\mathrm{\dot{m}_{c}}=1.9 \times 10^{-14}$ M$_\odot$/year, indicating that it is not possible to ablate the companion within the Hubble time scale. 

 \subsection{Origin of DM variations}
In this paper, we report time-dependent as well as frequency-dependent DM variations for PSR J1544+4937. The time-dependent DM variations observed for this pulsar is in the same range (order of $10^{-3} \mathrm{cm^{-3}}$) as observed for other BW MSPs (e.g. PSR J2051$-$0827, \cite{shaifullah201621} and PSR J2214+3000, \cite{bak2020timing}). The reason for the time-dependent DM variations could be the motion of the pulsar or the dynamical evolution of the ISM. 

We also probed the origin of frequency-dependent DM variations. There could be two major causes for the frequency-dependent DM variation \citep{donner2019first}. Firstly, equation (\ref{eqn:dispersion law}) is only valid under certain conditions ($\mathrm{f_{p}=8.98 \, kHz \times \sqrt{\frac{n_{e}}{cm^{-3}}}<< f}$ and $\mathrm{f_{c}=2.80 \, MHz \times\frac{B}{1G}<<f}$, where $f_{p}$ is the plasma frequency, $f_{c}$ is the cyclotron frequency and $f$ is the observing frequency). The use of equation (\ref{eqn:dispersion law}) under extreme conditions (such as low frequencies, large electron density, or high magnetic field strength) could give rise to frequency-dependent DM. First order deviation from equation (\ref{eqn:dispersion law}) is given as \citep{donner2019first},

\begin{equation}
   \mathrm{ t_{1}-t_{2}= D \times  DM(\frac{1}{f_{1}^{2}}-\frac{1}{f_{2}^{2}})(1+T_{1}+T_{2})}
\end{equation}
where $\mathrm{t_{1}}$ and $\mathrm{t_{2}}$ are time delays at the observing frequencies $\mathrm{f_{1}}$ and $\mathrm{f_{2}}$. With,
\begin{equation}
\mathrm{T_{1}=3f_{p}^{2}(f_{1}^{2}+f_{2}^{2})/4f_{1}^{2}f_{2}^{2}}
\label{eqn:T1}
\end{equation}
and
\begin{equation}
\mathrm{T_{2}=\pm2f_{c}\cos \gamma(f_{2}^{3}-f_{1}^{3})/(f_{1}^{2}-f_{2}^{2})f_{1}f_{2}}
\label{eqn:T2}
\end{equation}
where $\gamma$ is the line of sight angle with the magnetic field.

The second reason for the frequency-dependent DM variation could be the spatial electron density variations in the ISM, which causes frequency-dependent refraction of radiation on different spatial scales. As a result of it, rays follow a random walk and different frequency rays cover different parts of ISM, sampling regions with different electron densities \citep{cordes2016frequency}. 

If we consider, that the frequency-dependent DM variations arise from the invalidity of equation (\ref{eqn:dispersion law}), then a DM difference between band 3 and band 4 of $\sim$ $4 \times 10^{-3}$ (from Figure \ref{Figure:FrequencydependentDM}), requires $\mathrm{T_{1}+T_{2}} \sim 3 \times 10^{-4} $. Using equations (\ref{eqn:T1}) and (\ref{eqn:T2}), we estimate values of $\mathrm{n_{e}}$ and $\mathrm{B_{||}}$ of the order of $\mathrm{10^{4}~ cm^{-3}}$ and $10^{-3}$ G. Such high electron density within the galaxy is not feasible considering the Galactic electron density models \citep[e.g.][]{yao2017new}. \cite{donner2019first} also argued that such a scenario is highly unfeasible for PSR J2219+4754 and favored the 2nd cause for the observed frequency-dependent DM.

\subsection{Origin of orbital variability}
We report a maximum and minimum $\mathrm{\dot{P}_{b}}$ detection of $7 \times 10^{-12}~\mathrm{s/s}$ and  $-5 \times 10^{-12}~\mathrm{s/s}$ for PSR J1544+4937. The orbital variability observed for PSR J1544+4937 from Figure \ref{Figure:To_variation} is smaller compared to the other spider MSP systems where such studies exist. For example, PSR J2051$-$0827 \citep{shaifullah201621}; 47 Tuc J, 47 Tuc O \citep{freire2017long}; 47 Tuc W \citep{ridolfi2016long} has $\mathrm{\Delta P_{b}}$ of the order of $10^{-8}$ days and for 47 Tuc V has $\mathrm{\Delta P_{b}}$ is of the order of $10^{-5}$ days \citep{ridolfi2016long}. 

However, orbital period variations are not observed for all spider MSPs. There are few such systems where no orbital period variations are observed, for example; PSR J2214+3000, PSR J2234+0944 \citep{bak2020timing}; PSR J0610$-$2100 \citep{desvignes2016high}; 47 Tuc I, 47 Tuc R \citep{freire2017long}. Therefore we need long-term timing studies of a larger sample of spider MSPs to realize why a set of such systems show orbital period variations whereas others do not.

Variation of the orbital period of the BW MSP systems can be caused by several effects as studied by \cite{lazaridis2011evidence} and \cite{doroshenko2001orbital}. In this section, we have followed the same treatment to probe the possible contributors to the observed variation of the orbital period for PSR J1544+4937. The major contributions are as follows,
\begin{equation}
\label{eqn:Pbdot}
\mathrm{\dot{P}_{b} = \dot{P}_{b}^{GW} + \dot{P}_{b}^{D} +\dot{P}_{b}^{\dot{m}} +\dot{P}_{b}^{Q}}
\end{equation}
where $\mathrm{\dot{P}_{b}^{GW}}$ is the contribution from the emission of gravitational radiation, $\mathrm{\dot{P}_{b}^{D}}$ is the doppler correction which is the combined effect of the proper motion and acceleration of the binary system, $\mathrm{\dot{P}_{b}^{\dot{m}}}$ is the contribution by the mass loss from the binary system ($\mathrm{\dot{m}}$) and $\mathrm{\dot{P}_{b}^{Q}}$ is the input from the variation of the gravitational quadrupole moment of the companion star.

In general relativity for circular orbits, the orbital decay due to gravitational wave emission is given by,
\begin{equation}
\mathrm{\dot{P}_{b}^{GW}=-\frac{192\pi}{5}\left(\frac{2\pi Gm_{c}}{P_{b}c^{3}}\right)^{\frac{5}{3}}\frac{q}{(q+1)^{\frac{1}{3}}}}
\end{equation}
where $q$ is the ratio of the pulsar mass ($\mathrm{m_{p}=1.4 M_\odot}$) to the companion mass ($\mathrm{m_{c}=0.017 M_\odot}$). This gives q=82. Thus, 
\begin{equation}
\mathrm{\dot{P}_{b}^{GW} = -1.73 \times 10^{-14}}
\end{equation}
The above value of the $\mathrm{\dot{P}_{b}^{GW}}$ is two orders of magnitude less than the observed orbital period variation.

The contribution from the second term in equation (\ref{eqn:Pbdot}) is the combined effect of two terms 
\begin{equation}
    \mathrm{\dot{P}_{b}^{D} = \dot{P}_{b}^{Shk} + \dot{P}_{b}^{acc}}
\end{equation}
where
\begin{equation}
 \mathrm{\dot{P}_{b}^{Shk}=\frac{(\mu_{\alpha}^{2}+\mu_{\beta}^{2})d}{c}P_{b}}
\end{equation}
$\mathrm{\mu_{\alpha}}$, $\mathrm{\mu_{\beta}}$ are the values of transverse proper motion in RA, DEC respectively, and d is the distance to the pulsar. This gives,
\begin{equation}
\mathrm{\dot{P}_{b}^{Shk} = 1.56 \times 10^{-14}} 
\end{equation}
The above value of the $\mathrm{\dot{P}_{b}^{Shk}}$ is also one order of magnitude less than the observed value of $\mathrm{\dot{P}_{b}}$.

Moreover, acceleration of the binary system can be caused by the presence of a third massive body near it or by the differential rotation of the galaxy. This acceleration has similar effects on both spin period derivative $\mathrm{\dot{P}}$ and orbital period derivative $\mathrm{\dot{P}_{b}}$. Therefore, assuming that the whole of the $\mathrm{\dot{P}}$ is as a result of acceleration, the maximum contribution from this term is, 
$\mathrm{(\frac{\dot{P}_{b}}{P_{b}})^{acc}}$ = $\mathrm{(\frac{\dot{P}}{P})^{acc}=1.2 \times 10^{-18} s^{-1}}$ , which is 5 orders of magnitude less than the observed variation.

Mass loss from the binary system can be another factor responsible for the observed $\mathrm{\dot{P}_{b}}$. Assuming a circular orbit, no mass-loss from the pulsar ($\mathrm{\dot{m}_{p}=0}$) and a mass loss rate of the companion to be $\mathrm{\dot{m}_{c}} $, we have,
\begin{equation}
    \mathrm{\left(\frac{\dot{P}_{b}}{P_{b}} \right)^{\dot{m}_{c}} = \frac{-2\dot{m}_{c}}{M}}
\end{equation}
where M = 1.417 M$_\odot$ is the total mass of the system ($\mathrm{m_{c} + m_{p}}$, Table \ref{tab:Table2}); $\mathrm{\dot{m}_{c} = 1.94 \times 10^{-14}}$ M$_\odot$ $\mathrm{year^{-1}}$ , we get,
\begin{equation}
   \mathrm{(\dot{P_{b}})^{\dot{m}_{c}}= 9 \times 10^{-18} }
\end{equation}
This contribution is 5 orders of magnitude less than the observed $\mathrm{\dot{P}_{b}}$. As the contribution from the first three terms in equation (\ref{eqn:Pbdot}) is much smaller than the observed variation of the orbital period, we conclude that it must arise from the last term.
Thus, the variation of the gravitational quadrupole moment of the companion is most likely to be the major cause for the observed orbital period variation in PSR J1544+4937. The same cause is also reported to be responsible for other BW MSPs, where the variation of the orbital period is seen, for example, PSR J2051$-$0827 \citep{shaifullah201621} and PSR J1959+2048 \citep{applegate1994orbital}.

In the following, we have applied the gravitational quadrupole coupling (GQC) model presented by the \cite{applegate1994orbital} to PSR J1544+4937 to explain the observed orbital period variation. The change in the orbital period corresponding to a variation of the quadrupole moment of the companion $\mathrm{\Delta Q}$, is given as \citep{applegate1994orbital},

\begin{equation}
    \mathrm{\left(\frac{\Delta P_{b}}{P_{b}}\right)^{Q} = -9\frac{\Delta Q}{m_{c}a^{2}}}
    \label{Eqns1}
\end{equation}
where a is the relative semi-major axis of the orbit ($\mathrm{a = a_{p}+a_{c}}$, $\mathrm{a_{p}}$ and $\mathrm{a_{c}= \frac{a_{p}m_{p}}{m_{c}}}$ are the semi-major axis of the orbit of the pulsar and the companion respectively).
The variation in quadrupole moment due to cyclic spin up and spin-down of a thin shell of radius; $\mathrm{R_{c}}$ and mass; $\mathrm{M_{s}}$ rotating with angular velocity; $\Omega$ in the gravitational field of a point mass; $\mathrm{m_{c}}$ is given by \citep{applegate1994orbital},
\begin{equation}
    \mathrm{\Delta Q = \frac{2}{9} \frac{M_{s} R_{s}^{5}}{Gm_{c}}\Omega \Delta \Omega}
    \label{Eqns2}
\end{equation}
where G is the Newtonian gravitational constant and $\mathrm{\Delta \Omega}$ is the change in the angular velocity of the shell. Combining equation (\ref{Eqns1}) and equation (\ref{Eqns2}) we get the change in the angular velocity resulting from the change in the orbital period and is given by,
\begin{equation}
    \mathrm{\frac{M_{s}}{m_{c}} \frac{\Delta \Omega}{\Omega} = \frac{G m_{c}}{2R_{c}^{3}}\left(\frac{a}{R_{c}}\right)^{2} \left(\frac{P_{b}}{2 \pi}\right)^2 \frac{\Delta P_{b}}{P_{b}}}
    \label{Eqns3}
\end{equation}
This change in angular velocity gives rise to a change in the luminosity,
\begin{equation}
\mathrm{\Delta L = \frac{\pi}{3} \frac{Gm_{c}^{2}}{R_{c}P_{mod}} \left(\frac{a}{R_{c}}\right)^{2} \frac{\Omega_{dr}}{\Omega} \frac{\Delta P_{b}}{P_{b}} }
\label{Eqns4}
\end{equation}
where $\mathrm{\Omega_{dr}}$ is the angular velocity of the differential rotation and $\mathrm{P_{mod}}$ is the period of the orbital period modulation. 

We have calculated the fractional change in angular velocity given by equation (\ref{Eqns3}) and the corresponding change in luminosity given by equation (\ref{Eqns4}) for PSR J1544+4937. For these calculations, we have used the values of the parameters as,
$\mathrm{R_{c}} = 0.085$ R$_\odot$ \citep[from optical study considering hot spot model and distance to the pulsar of 2.4 kpc,][]{tang2014identification},
$\mathrm{m_{p}=1.4}$ M$_\odot$, $\mathrm{m_{c} = 0.017 M_\odot}$,
$\mathrm{P_{mod}} = 5.8$ years (inferred from Figure \ref{Figure:To_variation}(b)),
a = $9 \times 10^{8}$ m \citep[calculated using the inclination angle = $52^{\circ}$,][]{tang2014identification},
$\mathrm{\frac{\Delta P_{b}}{P_{b}} = 3.7 \times 10^{-8}}$ (inferred from Figure \ref{Figure:To_variation}(b)). Thus we get,
\begin{equation}
    \mathrm{\frac{\Delta \Omega}{\Omega} = 1.2 \times 10^{-3}}
\end{equation}
where we have assumed $\mathrm{M_{s} = 0.1~m_{c}}$, as assumed by \cite{applegate1994orbital} in their model. From equation (\ref{Eqns4}) we get, 
\begin{equation}
   \mathrm{\Delta L = 7.8 \times 10^{29} erg/s}
\end{equation}
where \cite{applegate1994orbital} assumes in their model an angular velocity $\mathrm{\Omega_{dr} = \Delta \Omega}$; and luminosity variations at the level; $\mathrm{\frac{\Delta L}{L} = 0.1}$. Using similar assumptions for PSR J1544+4937 we get,
\begin{equation}
   \mathrm{L = 7.8 \times 10^{30} erg/s}
    \label{Eqn7}
\end{equation}

From optical observations of PSR J1544+4937, effective temperature is $\mathrm{T_{eff}}$ = 4300 K \citep{tang2014identification} and corresponding internal luminosity is L =$\mathrm{4\pi R_{c}^{2} \sigma T_{eff}^{4}  \sim 8.5 \times 10^{30}}$ erg/s . This luminosity from the optical study is comparable to the luminosity derived in equation (\ref{Eqn7}). Thus, the GQC model can explain the orbital period variations observed for PSR J1544$+$4937. This model also explained the orbital period variation observed for PSR B1957$+$20 \citep{applegate1994orbital}. \cite{lazaridis2011evidence} also applied this model to PSR J2051$-$0827 where they concluded that under specific assumptions the GQC model can explain the orbital period variations of PSR J2051$-$0827.

\subsection{Inclusion of PSR J1544+4937 in PTA}
Orbital parameter variability is commonly observed for the BW MSPs  \citep[e.g variation of orbital period and/or projected semi-major axis,][]{shaifullah201621,bak2020timing} along with DM variation over long time scales.
To account for these orbital period variations, we have added 5 additional higher-order orbital frequency derivatives in the BTX model for PSR J1544+4937.
The addition of more parameters to the timing model reduces the sensitivity to a gravitational wave (GW) signal as the parameters of the pulsars are determined from the same data set. But using stimulated data \cite{bochenek2015feasibility} showed that fitting with many higher order orbital frequency derivatives only absorbs less than 5$\%$ of the low-frequency spectrum expected from a stochastic GW background signal. This emphasizes that if the timing systematics in BW MSPs can be modeled precisely by adding higher frequency derivatives, then BW MSPs should also be included in PTA. For example, despite showing orbital period variability the BW MSP J0023+0923 is included in the European pulsar timing array \citep{bak2020timing}. With the increasing number of newly discovered pulsars in compact orbits, the addition of even a few BW MSPs can lead to significant improvement in PTA sensitivity towards GW detection.
\begin{figure}
\centering
\includegraphics[trim={0cm 0cm 0cm 0cm},clip,scale=0.28]{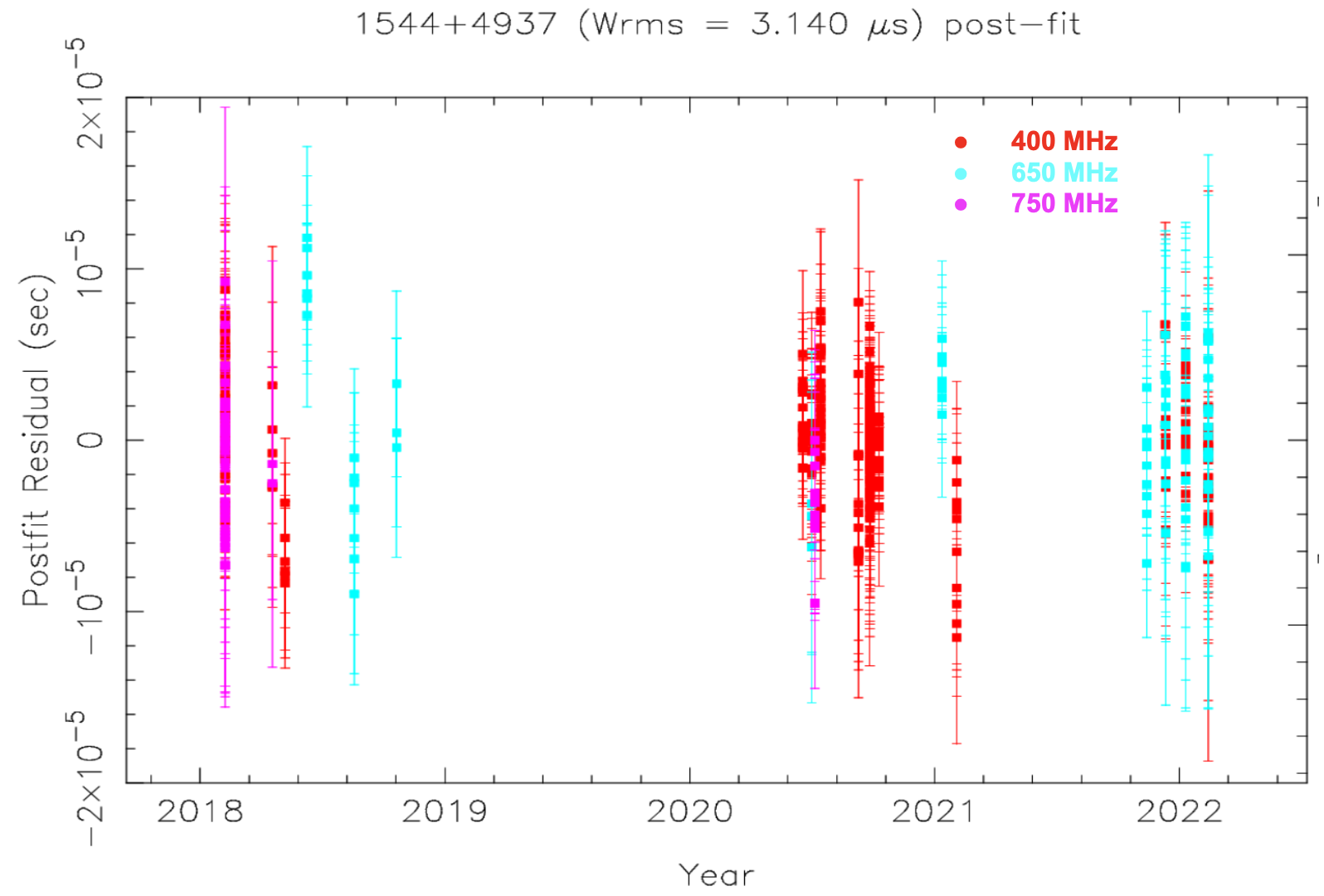}
\caption{Post-fit timing residual of 3.1 $\mu s$ for PSR J1544+4937 using uGMRT observations having 200 MHz bandwidth.} 
\label{Figure:uGMRT residual}
\end{figure}

The 5.5 $\mathrm{\mu s}$ timing residuals along with the remaining trends indicates that PSR J1544+4937 may not be an ideal MSP to include in the PTA in the current state of timing study.
However, we have got a timing residual of 3.1 $\mathrm{\mu s}$ while using only the uGMRT system having 200 MHz bandwidth from 2018$-$2022 (Figure \ref{Figure:uGMRT residual}). This timing precision is higher than the timing precision of PSR J2214+3000 ($\sim$ 4.3 $\mathrm{\mu s}$) which has already been added in the European pulsar timing array \citep{bak2020timing}. 
The ongoing observations with the uGMRT will allow us to add additional sensitive simultaneous dual frequency TOAs; which may enable better modeling of the PSR J1544+4937, increasing its possibility of inclusion in the PTA.

\vspace{0.5cm}
\setlength{\parindent}{0cm}
We acknowledge the support of the Department of Atomic Energy, Government of India, under project no.12-R\&D-TFR-5.02-0700. The GMRT is run by the National Centre for Radio Astrophysics of the Tata Institute of Fundamental Research, India. We acknowledge the support of GMRT telescope operators for observations. We are grateful to Scott Ransom for providing us with the GBT data for this pulsar. We want to thank Paul Ray for the help regarding folding of LAT photons. We want to thank Paulo Freire for his useful comments to study the orbital period variation for this pulsar. We thank the anonymous reviewer of the paper for insightful comments and suggestions.

\bibliography{J1544_timing_ver0.bbl}{}
\bibliographystyle{aasjournal}


\end{document}